\documentclass[aps,rmp,nofootinbib,12pt,tightenlines]{revtex4}

\usepackage{amsmath,amssymb}
\usepackage[mathcal]{euscript}
\usepackage{graphicx}
\usepackage{psfrag}

\usepackage{natbib}

\newcommand{\NS}{Navier--Stokes}
\newcommand{\FP}{Fokker--Planck}

\newcommand{\MB}{Maxwell--Boltzmann}

\newcommand{\mathnotation}[2]{\newcommand{#1}{\ensuremath{#2}}}
\newcommand{\Trless}[1]{\overset{\circ}{#1}}

\newcommand{\Dt}[1]{\frac{D#1}{D\time}}

\renewcommand{\l}{\left}                        
\renewcommand{\r}{\right}                       
\mathnotation{\pd}{\partial}                    
\mathnotation{\pdt}{\partial_\time}             
\mathnotation{\ee}{{\mathrm e}}                 
\mathnotation{\grad}{{\boldsymbol{\nabla}}}     
\mathnotation{\lapl}{\Delta}                    

\mathnotation{\curl}{\grad\times}               
\renewcommand{\div}{\grad\cdot}                 

\mathnotation{\dint}{\,{\mathrm{d}}}            
\mathnotation{\ldef}{\mathrel{\raisebox{.069ex}{:}\!\!=}}
\mathnotation{\rdef}{\mathrel{=\!\!\raisebox{.069ex}{:}}}
\mathnotation{\half}{\tfrac{1}{2}}              
\mathnotation{\threehalf}{\tfrac{3}{2}}         
\mathnotation{\pheq}{&\phantom{=}}              
\renewcommand{\time}{t}                         
\mathnotation{\x}{x}                            
\mathnotation{\xv}{{\mathbf{\x}}}               
\mathnotation{\vc}{v}                           
\mathnotation{\vv}{{\mathbf{\vc}}}              
\mathnotation{\solidang}{\Omega}                

\mathnotation{\uc}{u}                           
\mathnotation{\uv}{{\mathbf{\uc}}}              
\mathnotation{\cc}{c}                           
\mathnotation{\cv}{{\mathbf{\cc}}}              
\mathnotation{\ku}{\hat{k}}                     
\mathnotation{\kuv}{{\mathbf{\ku}}}             
\mathnotation{\expar}{\coltime}                 
\mathnotation{\visc}{\mu}
\mathnotation{\tcond}{\eta}                     
\mathnotation{\tdiff}{\kappa}                   
\mathnotation{\Liouv}{{\mathcal{D}}}            
\mathnotation{\Coll}{{\mathcal{C}}}             
\mathnotation{\LColl}{{\mathcal{L}}}            
\mathnotation{\LCollt}{\widetilde{\LColl}}      
\mathnotation{\accel}{a}                        
\mathnotation{\accelv}{{\mathbf{\accel}}}       
\mathnotation{\mass}{m}                         

\mathnotation{\dens}{\rho}                      
\mathnotation{\ndens}{n}                        
\mathnotation{\Temp}{T}                         
\mathnotation{\Entr}{S}                         
\mathnotation{\Pres}{P}                         
\mathnotation{\Prest}{{\mathbb{\Pres}}}         
\mathnotation{\pres}{p}                         
\mathnotation{\unitI}{{\mathbb{I}}}             
\mathnotation{\Qt}{Q}                           
\mathnotation{\Q}{q}                            
\mathnotation{\Qv}{{\mathbf{\Q}}}               
\mathnotation{\ros}{E}                          
\mathnotation{\rost}{\mathbb{\ros}}             
\mathnotation{\tros}{\Trless{\ros}}             
\mathnotation{\trost}{\Trless{\rost}}           
\mathnotation{\gasc}{R}                         
\mathnotation{\Cv}{C_v}                         
\mathnotation{\Pra}{\sigma}                     
\mathnotation{\DtS}{\Sigma}
\mathnotation{\divPp}{\Psi}
\mathnotation{\divPpv}{\boldsymbol{\divPp}}

\mathnotation{\velT}{\varpi}                    
\mathnotation{\tRTi}{a}                         
\mathnotation{\Afp}{A}                          
\mathnotation{\Bfp}{B}                          
\mathnotation{\coltime}{\tau}                   
\mathnotation{\fdist}{f}                        
\mathnotation{\fdistt}{\tilde{f}}               
\mathnotation{\inv}{\psi}                       
\mathnotation{\Kern}{\mathcal{K}}               
\mathnotation{\Sourcef}{\mathcal{J}}
\mathnotation{\ccz}{\mathfrak{a}}
\mathnotation{\cci}{\mathfrak{b}}
\mathnotation{\ccii}{\mathfrak{c}}
\mathnotation{\cciii}{\mathfrak{d}}

\mathnotation{\Prandtl}{\mathrm{Pr}}            
\mathnotation{\pdegree}{d}                      

\begin{document}

\title{Continuum Equations for Stellar Dynamics}
\thanks{Published in \emph{Stellar Astrophysical Fluid Dynamics: Proceedings
of the Chateau de Mons meeting in honour of Douglas Gough's 60th birthday}
(Cambridge University Press, 2003).}

\author{Edward A. Spiegel}
\email{eas@astro.columbia.edu}
\affiliation{Department of Astronomy, Columbia University, New York, NY 10027}
\author{Jean-Luc Thiffeault}
\email{jeanluc@mailaps.org}
\affiliation{Department of Applied Physics and Applied Mathematics,
Columbia University, New York, NY 10027}


\begin{abstract}

The description of a stellar system as a continuous fluid represents a
convenient first approximation to stellar dynamics, and its derivation from
the kinetic theory is standard.  The challenge lies in providing adequate
closure approximations for the higher-order moments of the phase-space density
function that appear in the fluid dynamical equations.  Such closure
approximations may be found using representations of the phase-space density
as embodied in the kinetic theory. In the classic approach of Chapman and
Enskog, one is led to the \NS\ equations, which are known to be inaccurate
when the mean free paths of particles are long, as they are in many stellar
systems.  To improve on the fluid description, we derive here a modified
closure relation using a \FP\ collision operator.  To illustrate the nature of
our approximation, we apply it to the study of gravitational instability.  The
instability proceeds in a qualitative manner as given by the \NS\ equations
but, in our description, the damped modes are considerably closer to
marginality, especially at small scales.

\end{abstract}

\maketitle

\section{A Kinetic Equation}

If we have a system of $N$ stars, with $N$ very large, and wish to study its
large-scale dynamics, we have to choose the level of detail we can profitably
treat.  Even if we could know the positions and velocities of all $N$
stars for all times, we would be mainly interested in the global properties
that are implied by this information.  For just such reasons, many
investigators prefer to find an approach that leads directly to a macroscopic
description of the dynamics.  As \textcite{Ogorodnikov} has put it, ``In order
to exhibit more clearly the kinematics of highly rarefied media, and of
stellar systems in particular, it is useful to make a comparison with the
motion of a fluid.''  However, traditional methods for deriving fluid
equations are effective only for media in which the mean free paths of
constituent particles are very short compared to all macroscopic scales of
interest.  This condition is not met in many stellar systems and plasmas and
so we here describe an approach that is effective for deriving fluid equations
for rarefied media such as stellar systems.

The first problem we must face is to decide what the kinetic description
of a stellar dynamical system ought to be.  Since an $N$-body
description is not what we want to work with, even if we could, since
that approach would have us computing the complicated trajectory of the
stellar system through a phase space with large dimension.  So we go
straight to the description of the system in the six-dimensional phase
space whose coordinates are the spatial coordinates ($x^i$) and the
velocities ($v^i$) of the $N$ stars, where $i=1,2,3$.

A plot of the locations of each of the $N$ stars in the six-dimensional
phase space at some given time, would reveal a swarm of
points whose detailed description would also be too complicated for us,
at least in a first look at the problem.  So instead, we concentrate on an
ensemble mean of such a description and seek an equation for the density
distribution of this mean.  That equation, on which we base this work, is
an evolution or continuity equation for the probability density
in the six-dimensional phase space.

Since we treat the stars as points, the true density $F$ in the
six-dimensional phase space is a summation of delta functions at
suitable locations.  This density is advected by a six-dimensional
phase velocity that Hamilton's equations tell us is solenoidal.
Hence the total time derivative of the density is \begin{equation}
        \Liouv F = 0
        \label{eq:liouv}
\end{equation}
where the comoving derivative in phase space is defined as
\begin{equation}
        \Liouv \ldef \pd_\time + {\bf v}\cdot\nabla_{\bf x} +
{\bf a}\cdot\nabla_{\bf v}
\ \label{eq:liouville}
\end{equation}
and the subscripts ${\bf x}$ and ${\bf v}$ on the gradient
symbols indicate that they are gradients with respect to position
and velocity respectively.  As usual, ${\bf v}=\dot {\bf x}$ where
the dot means total time derivative and the quantity ${\bf a}$ is the
gravitational acceleration per unit mass; we assume that all the stars
in the system have the same mass, $m$.  The gravitational
acceleration is given by the gradient of the gravitational
potential per unit mass, which is a solution of Poisson's
equation, \begin{equation}
        \Delta \Phi = 4\pi\, G \int F\, \dint^3\vc.
        \label{eq:poisson}
\end{equation}

Let $f({\bf x}, {\bf v}, t)$ be the ensemble mean of $F$.  Then the total
density will be $F=f+\hat f$ where $\hat f$ represents the fluctuations about
the ensemble mean; $\hat f$ will involve the same summation of delta functions
as does $F$ plus a smooth background distribution with negative mass density
arranged so that the ensemble average of $\hat f$ is zero.  We similarly split
the gravitational potential $\Phi$ into an ensemble mean part $\phi$ plus a
fluctuating part $\skew{6}\hat \phi$.  Then, if we take the ensemble average
of (\ref{eq:liouv}), we obtain
\begin{equation}
        \partial_t f + {\bf v}\cdot \nabla_{\bf x} f +
        (\nabla_{\bf x} \phi) \cdot \nabla_{\bf v} f =
        -\langle (\nabla_{\bf x}\skew{6}\hat \phi) \cdot \nabla_{\bf v} \hat
        f\rangle    \ ,
        \label{eq:kineq}
\end{equation}
where 
\begin{equation}
        \Delta \phi = 4\pi G \int f \dint^3\vc.
\end{equation}

The terms on the left hand side of (\ref{eq:kineq}) describe how the mean
phase density, or distribution function, $f$, streams through the
single-particle phase space.  The right side represents the mean influence on
the evolution of $f$ exerted by the average of the fluctuation-interaction
term. The latter represents the self-interactions of the system caused by
fluctuating effects and may be thought of as representing the influence of
collective modes (such as waves or quasiparticles) on the motions of the
individual particles.

It is typical that the chance of close approaches of two stars in many stellar
systems is small.  Because of this, one commonly made approximation is to
neglect the right side of (\ref{eq:kineq}) completely and so work with what is
called the collisionless Boltzmann (or Vlasov) equation.  More reasonably
perhaps, one may try to give an expression for the way the
self-interaction term affects the flow of the phase density through the phase
space.

As in Boltzmann theory, we shall suppose that the right side
of (\ref{eq:kineq}) may be expressed as a functional of $f$ itself so that
the kinetic equation is deterministic.  That is, we assume that the
kinetic equation takes the form \begin{equation}
        \partial_t f + {\bf v}\cdot \nabla_{\bf x} f
        + (\nabla_{\bf x} \phi)\cdot \nabla_{\bf v} f = \Coll[\fdist]
        \label{eq:Boltzmann}
\end{equation}
where $ \Coll[\cdot]$ may be called a collision term in keeping with the
terminology of kinetic theory even though it does not arise from direct
binary collisions.  This approach may be acceptable because it appears
that the fluid description that we seek is not very sensitive to the
details of the right side of (\ref{eq:kineq}).  On the other hand, we must
admit that this hope is founded on a very limited range of trial forms
since the job of deriving the consequences of each form is laborious.
Our aim here is to adopt one standard form for the collision term and use
it to go on to a coarser description of the stellar system like that of
fluid dynamics.

The parallel to the Boltzmann theory has been used to good effect in the study
of plasma physics, as in the work of \textcite{Rosenbluth1957}.  As
\textcite{ClemmowDougherty} explain, those authors obtained their results by
``expanding the Boltzmann collision operator under the approximation that all
the deflections are small angle and cutting off the impact parameter at about
the Debye length\dots . At first sight the success of that method is
surprising, as any treatment dealing with binary collisions would seem to be
discredited.  The physical reason for the agreement is that, for the majority
of particles, there is little difference between a succession of numerous
small-angle collisions (regarded as instantaneous and occurring at random) and
the stochastic deflections due to the presence of many nearby particles
continually exerting weak forces.  These two pictures of the dynamics are of
course represented respectively by the Boltzmann and the \FP'' approaches.

Similar thoughts have been expressed in the context of stellar dynamics, most
recently by \textcite{Griv2001} and by kinetic theorists generally.  As E. G.
D. Cohen reports (\citeyear{Cohen1997}), ``when Academician Bogolubov and I
discussed the nature of kinetic equations, he mentioned a discussion he had
had with Professor A. Vlasov, where they had agreed that: {\it Yes, in first
approximation the kinetic equations for gases with strong short-range forces
(i.e. the Boltzmann equation) and for gases with long-range forces (i.e. the
Vlasov equation) differ, but in higher approximations they will become more
and more similar.}  How right they were.''  A formal theory to buttress these
remarks would be very comforting, but though we do not have one we shall adopt
the point of view that the Fokker--Plank terms capture the essence of
interactions of the stars in the system.  Having thus supported our approach
by the appeal to authority, we turn to the main purpose of this work, the
derivation of continuum mechanics from the microscopic theory in a way that is
not severely restricted to the case of short mean free paths.

\section{The Collision Term}

In the spirit of standard kinetic theories, we shall suppose that the effect
of the collision term is to drive the system toward a local equilibrium,
though the correct equilibrium of a stellar system is not known on purely
theoretical grounds.  The tendency to approach an equilibrium seems not even
to require a collision term of the usual kind since the violent relaxation
described by Lynden-Bell apparently can do the job.  Nevertheless, we shall
proceed in terms of the kinetic theory under the assumption that the spreading
of the phase fluid through the phase space may be effected by a collision
term.  Furthermore, though the gravitational force has long range in physical
space, we shall presume that this spreading takes the form of a diffusion of
$f$ through velocity space, that is, by the agency of a Fokker--Planck form of
the collision term.  (This may not be completely unfounded since there seems
to exist a form of gravitational shielding \cite{SpiegelSchucking} that may
support the idea of local behaviour.) In this spirit, we write
\begin{equation}
        \Coll[\fdist] = \frac{\pd}{\pd\vc^i}\l[
        -\Afp^{i} \fdist + \tfrac{1}{2}\,\frac{\pd}{\pd\vc^j}
        \l(\Bfp^{ij}\, \fdist\r)\r].
        \label{eq:FP}
\end{equation}
The coefficients $\Afp^{i}$ and $\Bfp^{ij}$ are generally functions of
$(\xv,\vv ,\time)$ and they may also be functionals of $\fdist$.  In this
discussion, the choice of these coefficients in the Fokker--Planck description
is adapted to the equilibrium that is expected or assumed.  This equilibrium
is a local one that satisfies the condition~\hbox{$\Coll[f_0]=0$}.

Our goal is to find equations that govern the dynamics of the
macroscopic properties of the fluid embodied in the density,
temperature and velocity. These are defined as:
\begin{alignat}{2}
&\text{Mass density}\qquad &\dens &\ldef \int\mass\fdist\dint^3\vc,
        \label{eq:massdensity}\\
&\text{Mean velocity}\qquad  &\uv &\ldef
        \frac{1}{\dens}\int \mass\vv\fdist\dint^3\vc,
        \label{eq:meanvelocity}\\
&\text{Temperature}\qquad  &\Temp &\ldef
        \frac{\mass}{3\gasc\,\dens}\int\cc^2\fdist\,\dint^3\vc,
        \label{eq:temperature}
\end{alignat}
where the \emph{peculiar velocity} is
\begin{equation}
        \cv(\xv,\vv,\time) \ldef \vv - \uv(\xv,\time),
\end{equation}
and~$R=k/m$, $k$ being Boltzmann's constant.

As to the nature of $\Coll[f]$, we shall design it so that it produces
what may be the simplest plausible equilibrium, namely the \MB\
distribution
\begin{equation}
        \fdist_0(\xv,\vv,\time) = \frac{\dens}{\mass(2\pi\gasc\Temp)^{3/2}}\,
                \exp\l(-\frac{\cc^2}{2\gasc\,\Temp}\r).
        \label{eq:MB}
\end{equation}
Since $T, \rho$ and ${\bf u}$ generally depend on ${\bf x}$ and $t$, this
is a local equilibrium and we choose \cite{ClemmowDougherty}
\begin{equation}
        \Afp^i = -\coltime^{-1}(\vc^i - \uc^i),\qquad
        \Bfp^{ij} = 2\coltime^{-1}\,\gasc\,\Temp\,\delta^{ij},
        \label{eq:FPcoeffs}
\end{equation}
so that~$\Coll[\fdist_0] = 0$.  We assume that the mean-free-time~$\expar$ is
a constant so that the \FP\ operator is linear in~$\fdist$.

The collision term adopted here ensures the conservation of mass,
momentum and energy in the system.  This is reflected in
the property
\begin{equation}
        \int \inv^\alpha\,\Coll[\fdist] \dint^3\vc = 0,\qquad
        \alpha=0,\ldots,4,
        \label{eq:conserv}
\end{equation}
where
\begin{equation}
        \inv^\alpha = \mass\,(1,\vv,\half\vc^2).
        \label{eq:collinv}
\end{equation}
Thus we neglect the possible effects of dissipative processes and of
evaporation of stars from the system.

The macroscopic quantities~\eqref{eq:massdensity}--\eqref{eq:temperature}
enter the equilibrium state~\eqref{eq:MB} about which we are expanding.
To ensure that the same macroscopic quantities that follow from $f$ are
those that determine $f_0$, we impose a consistency requirement known as
the \emph{matching conditions},
\begin{equation}
        \int\inv^\alpha\,\fdist\dint^3\vc
                = \int\inv^\alpha\,\fdist_0\dint^3\vc,\qquad
        \alpha = 0,\dots,4.
        \label{eq:matching}
\end{equation}

\section{Fluid Equations}
\label{sec:fluideq}

When we multiply the kinetic equation~\eqref{eq:Boltzmann} by
the collisional invariants \eqref{eq:collinv} and integrate over~$\vv$, the
right-hand side does not contribute to the outcome, and we are left with
\begin{align}
        \pd_\time\dens + \div(\dens\,\uv) &= 0,
        \label{eq:densfl}\\
        \pd_\time\uv + \uv\cdot\grad\uv &= -\dens^{-1}\div\Prest
                - \grad\phi,
        \label{eq:ufleq}\\
        \threehalf\dens\gasc\l(\pd_\time\Temp + \uv\cdot\grad\Temp\r)
                &= -\Prest:\grad\uv - \div\Qv,
        \label{eq:Tfleq}
\end{align}
where $\grad$ means~$\nabla_{\bf x}$.  Here the pressure
tensor~$\Prest$ and heat flux~$\Qv$ are defined as
\begin{equation}
        \Prest \ldef \int\mass\cv\cv\fdist\dint^3\vc,\qquad
        \Qv \ldef \int\half\mass\cc^2\cv\fdist\dint^3\vc.
        \label{eq:PresQdef}
\end{equation}
We see that the form of the macroscopic equations is just that of the
usual fluid equations.  This result is independent of the rarity of the
medium.  The usefulness of these equations depends entirely on how well
we can prescribe the higher-order moments~$\Prest$ and~$\Qv$.  A standard
way to proceed is to solve~\eqref{eq:Boltzmann} approximately
for~$\fdist$.  We shall follow this route also, but will deviate from the
normally used prescription at a certain point.

We let $\fdist = \fdist_0 + \expar\,\fdist_1 + \dots$ and look first at
order~$\expar^0$.  We find that~$\Coll[\fdist_0] = 0$, and the
solution~$\fdist_0$ is the \MB\ equilibrium~\eqref{eq:MB}.
From \eqref{eq:PresQdef}, we see that~\hbox{$\Prest_0=\pres\,\unitI$}
and~\hbox{$\Qv_0=0$}, where the scalar pressure is given
by \hbox{$\pres\ldef\dens\gasc\,\Temp$}. If we stop at this order,
Equations~\eqref{eq:densfl}--\eqref{eq:Tfleq} are then the Euler
equations for an ideal fluid.

At order~$\expar^1$, it is convenient to factor out the \MB\
solution from~$\fdist_1$ and write the equation to be solved as
\begin{equation}
        \LCollt\fdistt_1 = \Liouv\ln\fdist_0\,,
        \label{eq:fdistt1eq}
\end{equation}
where~\hbox{$\fdistt_1 \ldef \fdist_1/\fdist_0$} and
\begin{equation}
        \LCollt\fdistt \ldef
                - \cv\cdot\nabla_{\cv}\fdistt
                + \gasc\,\Temp\,\nabla_{\cv}\cdot(\nabla_{\cv}\fdistt).
        \label{eq:LColltdef}
\end{equation}
The right-hand side of~\eqref{eq:fdistt1eq} may be written out as
\begin{equation}
        \Liouv\l[-\frac{\cc^2}{2\gasc\,\Temp}
                + \ln\frac{\ndens}{(2\pi\gasc\Temp)^{3/2}}\r]
        = -\frac{1}{\gasc\,\Temp}\,\cv\cdot\Liouv\cv
                + \l(\frac{\cc^2}{2\gasc\,\Temp} - \frac{3}{2}\r)
                        \Liouv\ln\Temp
                + \Liouv\ln{\dens}.
        \label{eq:RHS}
\end{equation}
We note that
\begin{equation}
	\Liouv {\bf c} = -\grad\phi - D{\bf u}/Dt - {\bf c}\cdot
	\nabla {\bf u},\quad \text{where} \quad
	D/Dt \ldef \partial_t + {\bf u\cdot\nabla}.
\end{equation}

The operator~$\LCollt$ maps polynomials in~$\cv$ to polynomials of the
same degree so, given the form of (\ref{eq:RHS}), we may seek a
solution~$\fdistt_1$ to~\eqref{eq:fdistt1eq} as a cubic in~$\cv$.
We write
\begin{equation}
        \fdistt_1 = \ccz + \cci_i\,\cc^i + + \ccii_{ij}\,\cc^i\cc^j
                    + \cciii_{ijk}\,\cc^i\cc^j\cc^k,
        \label{eq:fdistt1cubic}
\end{equation}
where~$\ccz$, $\cci$, $\ccii$ and ~$\cciii$ are functions of~$\xv$
and~$\time$, and symmetric in their indices; repeated indices are summed.
Inserting~\eqref{eq:fdistt1cubic} into the left-hand side
of~\eqref{eq:fdistt1eq}, we obtain
\begin{equation}
        \LCollt\fdistt_1 = -3\cciii_{ijk}\,\cc^i\cc^j\cc^k
                - 2\ccii_{ij}\,\cc^i\cc^j
                + (6\gasc\,\Temp\cciii_{i\ell\ell} - \cci_i)\cc^i
                + 2\gasc\,\Temp\ccii_{\ell\ell}.
        \label{eq:LColltfdistt1}
\end{equation}
There is no~$\ccz$ term because it is annihilated by the \FP\ collision
operator.  We now equate coefficients of~$\cv$
between~\eqref{eq:LColltfdistt1} and~\eqref{eq:RHS}, and find
\begin{equation}
        \cciii_{ijk}
        = -\frac{1}{18\gasc\,\Temp}
        \l(\delta_{ij}\,\nabla_{\x^k}\ln{\Temp}
                + \delta_{ik}\,\nabla_{\x^j}\ln{\Temp}
                + \delta_{kj}\,\nabla_{\x^i}\ln{\Temp}\r),
        \label{eq:cciii}
\end{equation}
\begin{equation}
        \ccii_{ij}= -\frac{1}{4\gasc\,\Temp}
                \l[(\nabla_{\x^j}\uc_i + \nabla_{\x^i}\uc_j)
                        + \Dt{\ln{\Temp}}\,\delta_{ij}\r],
        \label{eq:ccii}
\end{equation}
\begin{equation}
        \cci_i
        = -\frac{1}{6}\,\nabla_{\x^i}\ln{\Temp}
                - \frac{1}{\gasc\,\Temp}\l(
		\nabla_{\x^i}\phi + \Dt{\uc_i}\r)
                - \nabla_{\x^i}\ln{\dens}.
        \label{eq:cci}
\end{equation}
Since~$\ccz$ is still unspecified, we may use it to satisfy the
matching condition~\eqref{eq:matching}
$\int\fdist_1\dint^3\vc = 0$, resulting from mass conservation.
Only the terms even in~$\cv$
contribute, and we find
\begin{equation}
        \int\fdist_1\dint^3\vc =
        \ndens(\gasc\,\Temp\,\ccii_{\ell\ell} + \ccz) = 0,
        \label{eq:densmatchfdist1}
\end{equation}
which allows us to solve for~$\ccz$  in terms of the trace of~$\ccii$.

The pressure tensor and heat flux~\eqref{eq:PresQdef} are then obtained
from~$\fdist_1$ by performing straightforward Gaussian integrals, and we get
\begin{equation}
        \Pres_{ij} = \pres\,\delta_{ij} + 2\pres\gasc\,\Temp\ccii_{ij},\qquad
        \Q_k = \half\pres\gasc\,\Temp\l(
                5\cci_k + 21\gasc\,\Temp\,\cciii_{k\ell\ell}
                \r).
        \label{eq:Q1}
\end{equation}
The~$\ccz$ term is absent from the pressure because we used the density
matching condition~\eqref{eq:densmatchfdist1}.  From~\eqref{eq:Q1}
and~\eqref{eq:ccii}, we find that the pressure tensor can be written
\begin{equation}
        \Prest = \pres\,\unitI - 2\visc\,\rost - \visc
                \l(\Dt{\ln{\Temp}} + \tfrac{2}{3}\,\div\uv\r)\unitI
        \label{eq:Pres1FP}
\end{equation}
to first order in~$\expar$,
where the viscosity~$\visc\ldef\half\pres\,\expar$, and
\begin{equation}
        \ros{}_{ij} \ldef \half\l(\nabla_{\x^j}\uc_i + \nabla_{\x^i}\uc_j
                        - \tfrac{2}{3}\,\div\uv\,\delta_{ij}\r)
        \label{eq:trostdef}
\end{equation}
is the rate-of-strain tensor in traceless form.

From~\eqref{eq:Q1},~\eqref{eq:cciii}, and~\eqref{eq:cci}, to first order
in~$\expar$, the heat flux is
\begin{equation}
        \Qv = -\tcond\,\grad{\Temp}
                - 3\tcond\,\Temp\l[
                \grad\ln{\pres} + \frac{1}{\gasc\,\Temp}\l(\Dt{\uv}
                + \grad\phi\r)\r]
        \label{eq:Qv1FP}
\end{equation}
where the thermal conductivity~$\tcond\ldef(5/6)\pres\,\expar\gasc$.

These results differ from those of the usual Navier--Stokes equations
for which $\Prest = \pres\,\unitI - 2\visc\,\rost$ and $\Qv =
-\tcond\,\grad{\Temp}$.  To get some understanding of the import of the
additional terms found here we introduce the specific entropy
\begin{equation}
        \Entr = \Cv\,\ln\bigl(\pres\,\dens^{-5/3}\bigr),
\end{equation}
where $\Cv\ldef 3\gasc/2$ is the specific heat at constant volume.
Since
\begin{equation}
\dot\Entr \ldef \Dt{\Entr} = \Cv\l[\Dt{\ln{\Temp}} +
\tfrac{2}{3}\, \div\uv\r] ,
\end{equation}
we find that
\begin{align}
        \Prest &= \pres\l(1 - \frac{\coltime}{2\Cv}\,\dot\Entr\r)\unitI
                - 2\visc\,\rost
                + \mathrm{O}(\coltime^2),\\
        \Qv &= -\tcond\,\grad\Temp + 3(\tcond\,\Temp/\pres)\div\mathbb{T}
                + \mathrm{O}(\coltime^2),
\end{align}
with $\mathbb{T} \ldef \l(\Prest - \pres\,\unitI\r)$.  If we put these results
into $\dot S$, we obtain
\begin{equation}
        \dot\Entr = -\frac{2\Cv}{3\pres}\,\l[\mathbb{T}:\grad\uv
                + \div\Qv\r] .
\end{equation}
Hence $\dot S$ can be seen to be $\mathrm{O}(\tau)$ and so the additional
terms in the pressure tensor are $\mathrm{O}(\tau^2)$.  A similar argument may
be made for the new terms in the heat flux.  Though these terms do not appear
in the conventional fluid equations, they can be quite significant when the
mean free paths are long.

If we eliminate the entropy using \begin{equation}
        \frac{1}{\Cv}\,\dot\Entr = \frac{\dot\pres}{\pres}
                - \frac{5}{3}\,\frac{\dot\dens}{\dens}
\end{equation}
we encounter the combination $p(t)-\tfrac{1}{2}\tau \dot p(t)$ which are
the first two terms of a Taylor series of $p$ in~$\tau/2$.  We may then
write
\begin{equation}
        \Prest = \pres(t - \tfrac{1}{2}\coltime)\,\unitI
                + \tfrac{10}{3}\,\visc\,\div\uv\,\unitI
                - 2\visc\,\rost
                        + \mathrm{O}(\coltime^2).
\end{equation}
We see that our procedure has taken account of the physical fact
that the medium senses what particles were doing one collision time
prior to the present time but is not yet aware of what they are doing at
the present.

\section{The Jeans Instability}

\newcommand{\transp}[1]{\widetilde{#1}}
\mathnotation{\third}{\tfrac{1}{3}}             
\mathnotation{\ddens}{\varphi}
\mathnotation{\dTemp}{\theta}
\mathnotation{\dpres}{\varpi}
\mathnotation{\Vgrav}{\phi}
\mathnotation{\Ggrav}{G}
\mathnotation{\densC}{\dens_{\Lambda}}
\mathnotation{\kJeans}{k_{\mathrm{J}}}
\mathnotation{\dkJeans}{k_{\mathrm{J}}}
\mathnotation{\apar}{a}
\mathnotation{\bpar}{b}
\mathnotation{\kvisc}{\nu}                      
\mathnotation{\cs}{a_\Temp}                     
\mathnotation{\css}{a_\Entr}                    
\mathnotation{\imi}{{\mathrm{i}}}               
\mathnotation{\grw}{\gamma}

As an application of the equations of
motion~\eqref{eq:densfl}--\eqref{eq:Tfleq}, we will use them together
with the pressure tensor and heat flux derived in
Section~\ref{sec:fluideq} to examine the Jeans criterion for
gravitational instability.  This instability, describing the
gravitational collapse of a homogeneous medium, was first investigated
by Jeans~(1929).  He found that perturbations above a critical
wavelength (the Jeans length) were unstable to gravitational collapse,
but that shorter wavelengths were unaffected due to the large-scale
nature of the gravitational force.  The Jeans length is the ratio of
the adiabatic sound speed~$\css$ to the gravitational
frequency~$\sqrt{4\pi\Ggrav\dens}$.  \textcite{Pacholczyk1959} and
\textcite{Kato1960} investigated the effect of viscosity and thermal
conductivity on the instability and found that the collapse occurs
above a critical wavelength given by the ratio of the isothermal sound
speed~$\cs$ to the gravitational frequency. That critical wavelength
is slightly larger than in the ideal case, which involves the
adiabatic sound speed.  Their interpretation for the increase in
critical wavelength is that temperature gradients are adverse to the
collapse, and a nonzero thermal conductivity allows the smoothing out
of these gradients through very slow displacements of the medium; the
mode is thus isothermal.

We will now investigate how the gravitational instability occurs in our set of
equations.  We take as our equilibrium a medium at rest and with uniform
density~$\dens_0$ and temperature~$\Temp_0$.  We expand each fluid variable
into an equilibrium piece and a small perturbation,
\begin{equation}
        \dens = \dens_0(1 + \ddens), \quad
        \Temp = \Temp_0(1 + \dTemp), \quad
        \pres = \dens\gasc\Temp = \pres_0(1 + \dpres), \quad
        \dpres = \ddens + \dTemp\,.
\end{equation}
The gravitational potential~$\Vgrav$ in~\eqref{eq:ufleq} is obtained from the
Poisson equation
\begin{equation}
        \lapl\Vgrav = 4\pi\Ggrav\,(\dens - \dens_0),
        \label{eq:Poissoneq}
\end{equation}
where we choose~$-\dens_0$ as a background ``neutralising'' density in order
to have a proper uniform equilibrium about which to expand; the equilibrium
velocity then vanishes.  The density~$-\dens_0$ is a repulsion term and may be
regarded as a Newtonian analogue of Einstein's cosmological constant.  To
leave it out as Jeans did and jump straight to the linearised
equation~\eqref{eq:Vgraveqlin} below is expedient but questionable.

The linearised equations of motion~\eqref{eq:densfl}--\eqref{eq:Tfleq} and
Poisson equation \eqref{eq:Poissoneq} are
\begin{gather}
        \pdt\ddens + \div\uv = 0,
                \label{eq:contlin}\\
        \dens_0\pdt\uv + \div\Prest =
                -\dens_0\grad\Vgrav,
                \label{eq:veleqlin0}\\
        \tfrac{3}{2}\pres_0\pdt\dTemp + \pres_0\div\uv
                + \div\Qv = 0,
                \label{eq:Tempeqlin}\\
        \lapl\Vgrav = 4\pi\Ggrav\dens_0\,\ddens.
        \label{eq:Vgraveqlin}
\end{gather}
The pressure tensor~$\Prest$ is given by~\eqref{eq:Pres1FP} and the heat
flux~$\Qv$ by~\eqref{eq:Qv1FP}.  We take the divergence of the velocity
equation~\eqref{eq:veleqlin0} and use the continuity
equation~\eqref{eq:contlin} to eliminate~$\div\uv$,
\begin{equation}
        \dens_0\pdt^2\ddens
                - \grad\grad:\Prest
        = 4\pi\Ggrav\dens_0^2\,\ddens,
        \label{eq:veleqlin}
\end{equation}
where we also used the Poisson equation~\eqref{eq:Vgraveqlin} to
eliminate~$\Vgrav$.  We then need to take two divergences of the linearised
pressure tensor,
\begin{equation}
        \grad\grad:\Prest =
                \pres_0\,\lapl(\ddens + \dTemp)
                + \visc\bigl(2\pdt\lapl\ddens
                - \pdt\lapl\dTemp\bigr).
        \label{eq:divdivP}
\end{equation}
We introduce the isothermal sound speed~$\cs$, the kinematic
viscosity~$\kvisc$, and the thermal diffusivity~$\tdiff$, through
\begin{equation}
        \cs^2 \ldef \frac{\pres_0}{\dens_0}, \qquad
        \kvisc \ldef \frac{\visc}{\dens_0}, \qquad
        \tdiff \ldef \frac{\tcond}{\dens_0\,C_p} =
                \frac{\tcond}{\tfrac{5}{2}\gasc\dens_0}.
        \label{eq:defs}
\end{equation}
Then, on inserting~\eqref{eq:divdivP} into~\eqref{eq:veleqlin}, we obtain
\begin{equation}
        \pdt^2\ddens =
        \cs^2\,\lapl(\ddens + \dTemp)
        + \kvisc\bigl(
                2\pdt\lapl\ddens - \pdt\lapl\dTemp\bigr)
        + \tfrac{5}{3}\kJeans^2\,\cs^2\,\ddens,
        \label{eq:densdisp}
\end{equation}
where the Jeans wavenumber is given by
\begin{equation}
        \kJeans^2 \ldef \frac{4\pi\Ggrav\dens_0}{\css^2}
                = \frac{3}{5}\,\frac{4\pi\Ggrav\dens_0}{\cs^2}\,.
\end{equation}
Next, we take the divergence of the linearised heat flux,
\begin{equation}
        \div\Qv = -\tcond\Temp_0\,\lapl\dTemp
                - 3\tcond\Temp_0\cs^{-2}
                \l[\pdt\div\uv + \lapl\Vgrav + \cs^2\lapl(\ddens + \dTemp)\r],
\end{equation}
and insert this into the temperature equation~\eqref{eq:Tempeqlin},
\begin{equation}
        \pdt\dTemp - \tfrac{2}{3}\pdt\ddens
                - \tfrac{5}{3}\tdiff\,\cs^{-2}
                        \l[4\cs^2\lapl\dTemp
                - 3\l(\pdt^2\ddens
                - \tfrac{5}{3}\kJeans^2\,\cs^2\,\ddens
                - \cs^2\lapl\ddens\r)\r] = 0,
        \label{eq:Tempdisp}
\end{equation}
where again we eliminated~$\div\uv$ using the continuity
equation~\eqref{eq:contlin}, and used the definition~\eqref{eq:defs} of the
thermal diffusivity~$\tdiff$.

\begin{figure}
\psfrag{k2}{{\large $k^2$}}
\psfrag{Reg}{{\large \ \ \ \ $\grw$}}
\centering
\includegraphics[width=.8\textwidth]{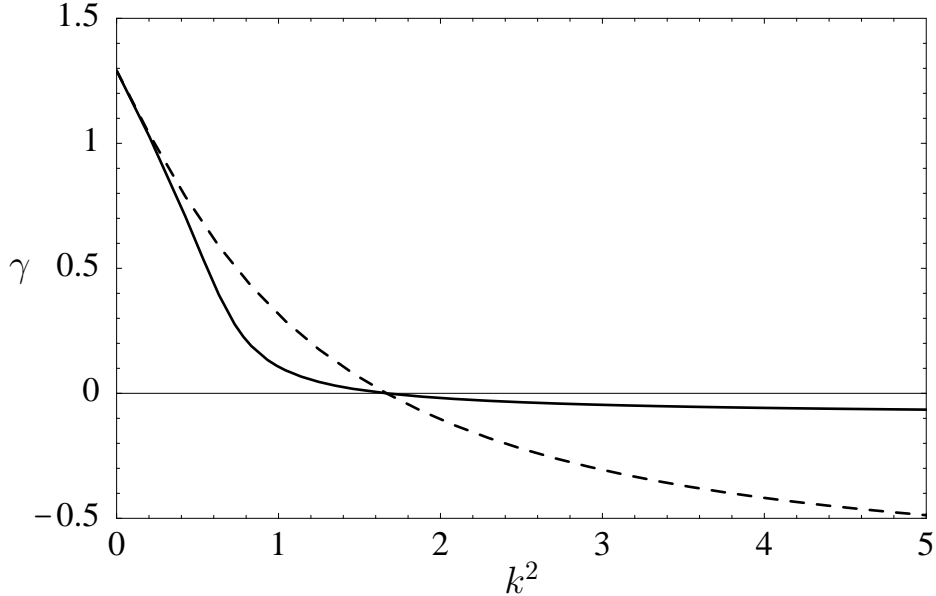}
\caption{Growth rate~$\grw$ as a function of wavenumber~$k$, for \NS\ (dashed
line) and the equations derived in Section~\ref{sec:fluideq} (solid).  The
growth rate of this mode is real and, in both cases, damping sets in above
the isothermal Jeans wavenumber~$(5/3)\dkJeans^2$ (here we have
taken~$\dkJeans=1$ in dimensionless units).}
\label{fig:directroot}
\end{figure}

Equations~\eqref{eq:densdisp} and~\eqref{eq:Tempdisp} are the equations
required to derive a dispersion relation for the gravitational
instability.  It is convenient to use the
viscous time~$\visc/\pres_0$ as unit of time and~$\cs\visc/\pres_0$ as
unit of length.
Recycling the same symbols for the dimensionless quantities turns
equations~\eqref{eq:densdisp} and~\eqref{eq:Tempdisp} into
\begin{gather}
        \l(\pdt^2 - \lapl - 2\pdt\lapl - \tfrac{5}{3}\dkJeans^2\r)\ddens
                + (\pdt - 1)\lapl\dTemp = 0,\\
        \l(3\pdt - \tfrac{40}{3}\lapl\r)\dTemp
                + \l(-2\pdt + 10
                \l(\pdt^2 - \tfrac{5}{3}\dkJeans^2
                - \lapl\r)\r)\ddens = 0.
\end{gather}
On letting~$\ddens, \dTemp \sim \exp(\imi k x + \grw t)$, we find the
dispersion relation
\begin{equation}
\l(\tfrac{3}{5} + 2 k^2\r)\grw^3
        + \tfrac{22}{15}\,k^2\,\grw^2
        + \l[\tfrac{22}{3}\,k^4
                + \l(1 - \tfrac{10}{3}\,\dkJeans^2\r)k^2 - \dkJeans^2\r]\grw
        + \tfrac{2}{3}\l(k^2 - \tfrac{5}{3}\,\dkJeans^2\r)k^2 = 0,
        \label{eq:disprelfull}
\end{equation}
which may be compared to the expression obtained from \NS\ by Kato \&
Kumar~(1960),
\begin{equation}
\tfrac{3}{5}\,\grw^3
        + \tfrac{22}{15}\,k^2\,\grw^2
        + \l(\tfrac{8}{9}\,k^4
                + k^2 - \dkJeans^2\r)\grw
        + \tfrac{2}{3}\l(k^2 - \tfrac{5}{3}\,\dkJeans^2\r)k^2 = 0,
        \label{eq:disprelNS}
\end{equation}
with the \FP\ values for the viscosity and thermal diffusivity inserted into
their result.  In each case the system is marginally stable with~$\grw=0$
at~\hbox{$k^2=(5/3)\dkJeans^2$}, and is damped for larger~$k$.  This is
illustrated in Fig.~\ref{fig:directroot}.
For~$k=0$, both dispersion relations predict a growth rate
of~$\grw(k=0)=\sqrt{5/3}\,\dkJeans$ in dimensionless units, which with
dimensions is~$\sqrt{4\pi\Ggrav\dens_0}$.  This is consistent with the fact
that dissipation is unimportant at large scales, so the growth rate at~$k=0$
involves only the gravitational time.

The asymptotic growth rate as~$k\rightarrow\infty$ is~$-3/4$ for \NS\ and
$-1/11$ for our system, independent of~$k$. (Multiply by~$\visc/\pres_0$
to recover dimensions.)  Thus, at large wavenumbers, the modes tend to be
uniformly damped, both in our case and for \NS.  This is because at large~$k$
the fluid behaves like a Stokes flow, where we can ignore the inertial and
gravitational terms completely, and the balance is between the Laplacian of
the pressure and the viscosity, which have the same number of spatial
derivatives; hence the lack of dependence on~$k$.  For our case (the equations
of Section~\ref{sec:fluideq}) there is a contribution from the new terms
that shifts the damping rate considerably closer to marginality.

\begin{figure}
\psfrag{k2}{\raisebox{-.5em}{\large $k^2$}}
\psfrag{Reg}{{\large $\mathrm{Re}\,\grw$}}
\centering
\includegraphics[width=.8\textwidth]{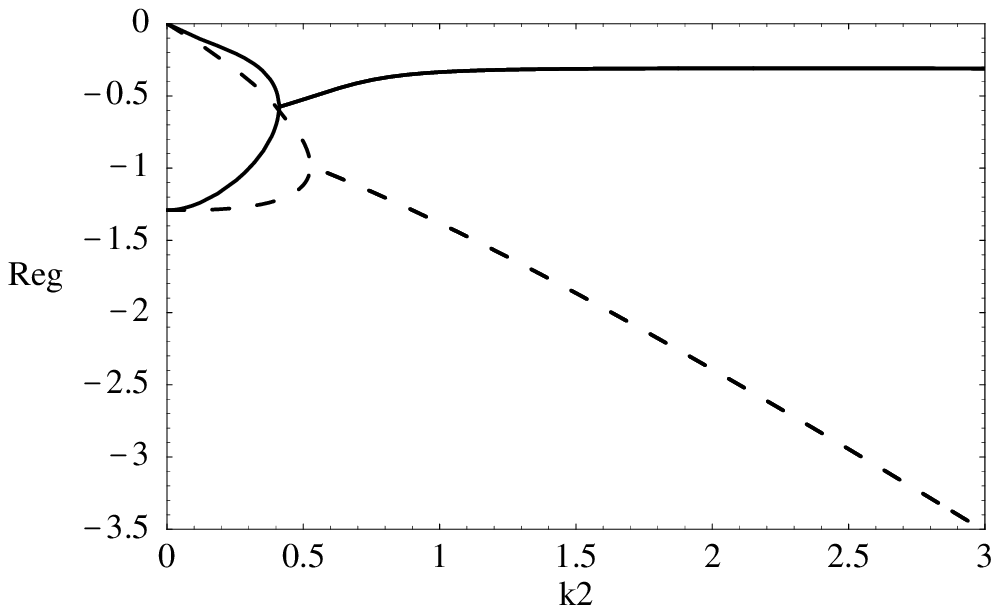}
\caption{Real part of the two complex roots~$\grw$ as a function of
wavenumber~$k$, for \NS\ (dashed line) and the equations derived in
Section~\ref{sec:fluideq} (solid).  Large wavenumbers are much more strongly
damped for \NS\ than in the system derived here.}
\label{fig:complexroot}
\end{figure}

Coexisting with the real root associated with the instability, there is also a
pair of unconditionally damped roots.  The real part of these roots is
plotted in Fig.~\ref{fig:complexroot}.
At~$k=0$ one of these roots is marginal (real part of growth rate equal to
zero) and is an \emph{isopycnal} mode (constant density).  However, this mode
is never destabilised and its growth rate immediately decreases as~$k$
increases away from zero.

For small~$k$ the two damped roots are distinct and real, but they come
together at larger~$k$ and become a complex-conjugate pair with nonzero
imaginary part (not plotted), indicating oscillatory behaviour.  The \NS\ case
(dashed line) is seen to have heavily damped complex roots at large
wavenumber.\footnote{In the \NS\ system the complex roots come together again
at very large~$k$, but the growth rate continues to decrease with~$k$.}  But
for our system of equation (solid line), the growth rate actually increases a
little, looking as though it may be headed for a Hopf bifurcation
(\textit{i.e.}, overstability, where the real part of the growth rate becomes
positive at nonzero imaginary part), before leveling off at an asymptotic
value of the damping rate given by~$-53/165\simeq-0.3212$.  Indeed, it can be
shown that there is no Hopf bifurcation for any realisable parameter values in
our equations, but the fact that the complex modes are somewhat
``destabilised'' by the new terms is intriguing (this is also true to a lesser
extent for the real mode described above).  This destabilisation has its
source in the~$k^2$ coefficient of the~$\grw^3$ term in the dispersion
relation~\eqref{eq:disprelfull}, which is not present in \NS, but its physical
significance is not yet apparent to us.

\section{Conclusion}

The basic approach in this as in other derivations of fluid equations from
kinetic theory is to write the general moment equations
(\ref{eq:densfl})--(\ref{eq:Tfleq}).  These are the fluid equations and, to
complete them, we need closure relations for the pressure tensor and heat
flux.  This is an issue astronomers are familiar with from the study of
radiative transfer.  For the purpose, we could invent a phenomenological
approximation as Eddington did in radiative transfer or we may pursue
approximate solutions of the kinetic theory as Hilbert did by expanding in the
collision time.  The Hilbert expansion was developed by Chapman and Enskog in
deriving the Navier--Stokes equations \cite{Uhlenbeck} and we have pursued
that line as well following earlier work \cite{ChenThesis,Chen2000,Chen2001a}
based on the relaxation model of kinetic theory
\cite{Bhatnager1954,Welander1954}.  However, in that latter work, as here, we
depart from the Chapman--Enskog approach in an essential way in not using
results from lower orders to to simplify the results in the current order.

To express this idea in equations, let us consider what happens in general
in such problems in the first order.  Once we have expressed the
one-particle distribution function as $f=f_0(1+\tau\varphi)$ where
$\tau$ is the (small) collision time, we are led to an equation for
$\varphi$ in the form \begin{equation}
{\cal L} \varphi = {\cal D}f_0 + {\cal O}(\tau) \label{first}
\end{equation}
where ${\cal L}$ is the linearisation of the collision operator.  In
general, ${\cal L}$ is self-adjoint, as it is in the present study.
Then ${\cal L}\psi^\alpha=0$ implies $\int \psi^{\alpha} {\cal L} \varphi
d{\bf v}=0$ and so we must have \begin{equation}
\int \psi^{\alpha} \left[{\cal D}f_0\right] d{\bf v} = {\cal O}(\tau).
\label{solve}
\end{equation}
Since $\psi^\alpha$ represents the collisionally invariant quantities,
the fluid equations to the current order may serve as the
solvability condition~(\ref{solve}).

In the Chapman--Enskog procedure, this solvability condition is taken to be a
lower order version of the fluid equations, here the Euler equation, and it is
used to simplify the right side of (\ref{first}).  Then, the results are used
in the general fluid equations.  For both of these two conditions to be
satisfied, we require $\tau$ to be very small indeed.

What we are doing here is to say that, to the first order, the fluid equations
themselves are a realisation of condition (\ref{solve}) and that it is
redundant to apply the same condition twice, once with $\mathrm{O}(\tau)$
retained and once with it omitted, as one does in the Chapman--Enskog method.
Rather, we simply use the full condition (\ref{solve}) as a compatibility
condition.  It is for this reason that, in the first order theory, we allow
ourselves to differ from the Chapman--Enskog results by terms
$\mathrm{O}(\tau^2)$.  In particular, we have for the trace of the pressure
tensor
\begin{equation}
  \mathrm{Tr}\ \Prest = 3\,\pres\l(1 - \frac{\coltime}{2\Cv}\,\dot\Entr\r)
                + \mathrm{O}(\coltime^2).
\end{equation}
This result differs from the exact trace ($3p$) by $\mathrm{O}(\tau^2)$ and
this, we suggest is allowed in a first order-theory.

We may add that in comparing the results of this approach to experiments on
ultrasound we find that they do better than the usual Navier--Stokes version.
Here we have an interesting example of a dictum of J. B. Keller: ``Two
theories may have the same accuracy but different domains of validity.''

We regret that though, in honour of Douglas Gough's birthday, we have gone to
second order in this approach (for only the relaxation model so far), we could
not fit the derivations into the space we were allotted in this volume.  So
those results will have to be presented elsewhere.  We are happy to report
that, in that next order, the trace of our pressure tensor differs from
the exact trace by $\mathrm{O}(\tau^3)$.  For now we must be content with
mentioning that result and presenting our best wishes to Douglas on his
birthday.


\end{document}